\documentclass[ba]{imsart}

\RequirePackage{amsthm,amsmath,amsfonts,amssymb}
\RequirePackage[numbers]{natbib}
\RequirePackage[colorlinks,citecolor=blue,urlcolor=blue,backref=page,backref=page]{hyperref}
\RequirePackage{graphicx}

\usepackage{amsmath}
\usepackage[ruled,vlined]{algorithm2e}

\newcommand{\Norm}{\mathcal{N}}
\newcommand{\bfY}{\textbf{Y}}
\newcommand{\bfeta}{\boldsymbol\eta}
\newcommand{\bfepsilon}{\boldsymbol\epsilon}
\newcommand{\bfyi}{\textbf{y}_i}
\newcommand{\bflambda}{\boldsymbol\lambda}

\lastpage{1}

\startlocaldefs
\theoremstyle{plain}

\theoremstyle{definition}

\theoremstyle{remark}

\endlocaldefs

\begin{document}

\begin{frontmatter}
\title{Sparse Bayesian Factor Models with Mass-Nonlocal Factor Scores}
\runtitle{Sparse Bayesian Factor Models with Mass-Nonlocal Factor Scores}

\begin{aug}
\author[1,A]{\fnms{Yingjie}~\snm{Huang}\ead[label=e1]{yingjie\_huang@brown.edu}},
\author[1,A]{\fnms{Dafne}~\snm{Zorzetto}\ead[label=e2]{dafne\_zorzetto@brown.edu}
}
\and
\author[A,B]{\fnms{Roberta}~\snm{De Vito}\ead[label=e3]{roberta\_devito@brown.edu}}
\address[1]{Equally contributing co-first authors.}
\address[A]{Data Science Institute, Brown University\printead[presep={,\ }]{e1,e2}}
\address[B]{Department of Biostatistics, Brow University\printead[presep={,\ }]{e3}}
\runauthor{Y. Huang et al.}
\end{aug}

\begin{abstract}
Bayesian factor models are widely used for dimensionality reduction and pattern discovery in high-dimensional datasets across diverse fields. These models typically focus on imposing priors on factor loading to induce sparsity and improve interpretability. However, factor scores, which play a critical role in individual-level associations with factors, have received less attention and are assumed to follow a standard normal distribution. This assumption oversimplifies the heterogeneity often observed in real-world applications.
We propose the sparse Bayesian Factor model with MAss-Nonlocal factor scores (BFMAN), a novel framework that addresses these limitations by introducing a mass-nonlocal prior on factor scores. This prior allows for both exact zeros and flexible, nonlocal behavior, capturing individual-level sparsity and heterogeneity. The sparsity in the score matrix enables a robust and novel approach to determine the optimal number of factors. Model parameters are estimated via a fast and efficient Gibbs sampler.
Extensive simulations demonstrate that BFMAN outperforms standard Bayesian factor models in factor recovery, sparsity detection, score estimation, and selection of the optimal number of factors. We apply BFMAN to the Hispanic Community Health Study/Study of Latinos, identifying meaningful dietary patterns and their associations with cardiovascular disease, showcasing the model's ability to uncover insights into complex nutritional data. 
\end{abstract}


\end{frontmatter}

\section{Introduction}

Bayesian factor models play a central role in numerous disciplines, including social sciences \citep{chen1992social}, genomics \citep{wang2011}, nutrition \citep{edefonti2012}, and more broadly in high-dimensional applications \citep{roy2021perturbed, casa2022parsimonious}. These models are particularly advantageous for large-scale data, providing a structured approach to dimensionality reduction, improving interpretability, and facilitating deeper understanding of the underlying data structure \citep{bernardo2003}. When dealing with high-dimensional datasets, incorporating sparsity or penalization techniques becomes critical for two primary reasons: first, to ensure interpretability and achieve meaningful insights into the data, and second, to guarantee that the covariance matrix is estimable \citep{lopes2004}.

Methodological developments have traditionally focused on imposing priors on the factor loading matrix, using approaches such as shrinkage priors \citep{bhattacharya2011sparse, legramanti2020bayesian}, sparsity priors \citep{carvalho2008, fruhwirth2024sparse}, spike-and-slab \citep{rovckova2016fast, bai2021spike}, and non-local mass priors \citep{Avalos2022HLDI}. However, little attention has been given to the factor score matrix, typically assumed to follow a standard multivariate normal distribution, implying independence between factors \citep{tucker1971relations, distefano2019understanding}. While some flexible approaches have been proposed using non-diagonal covariance structures for the factor scores, they often increase model complexity without addressing individual-level heterogeneity. \citep{mcdonald1967comparison}.

Factor scores play a critical role, as they quantify the score that each individual expresses on the corresponding factors, making them particularly relevant in various applications \citep{ de2021bayesian, castello2016association}. In applications such as nutritional epidemiology, factor analysis is often used to estimate both dietary patterns (i.e., factor loadings) and factor scores, to estimate the association between these patterns and disease outcome \citep{castello2016association}. Therefore, a more refined estimation and appropriate prior specification for factor scores are essential to accurately model the relationship between each factor and health outcomes. For instance, in diet-related studies, individuals may exhibit substantial heterogeneity in adherence to dietary patterns---some may strictly follow a given pattern, while others do not at all. Standard Gaussian assumptions fail to capture this heterogeneity and lack the flexibility to induce sparsity in individual-level scores.

To address these limitations, we introduce the Sparse Bayesian Factor Model with Mass-Nonlocal Factor Scores (BFMAN), a novel approach that assumes a mass-nonlocal prior directly on the latent factor scores. This framework introduces a more flexible posterior distribution for factor scores, characterizing the heterogeneity in subject-level associations with latent factors. The mass-nonlocal prior with a non-null probability allows for exact zero in the score matrix, and a non-local slab prior that do not overlap the spike yielding a sparse, heterogeneous structure that reflects real-world variation in individual behavior. 
 
Our model incorporates three key features enabled by this sparse prior on factor scores. First, sparsity in the score enhances interpretability by linking each latent factor to a small subset of individuals.  Second, when the sparsity assumption holds, it improves estimation accuracy and model efficiency. 
Third, inducing sparsity enables a novel, principled approach to inferring the number of latent factors. 
 While existing approaches typically focus on the sparsity or shrinkage in the factor loading matrix \citep{bhattacharya2011sparse, carvalho2008} or proportion of variance explained  \citep{de2021bayesian}, our method takes a novel approach by leveraging the level of sparsity in the factor score matrix to infer the optimal number of factors. 
 This unique perspective allows for more precise identification of factors and better captures the structural complexity of the data. 
To ensure computational scalability, we develop a fast and efficient Gibbs sampler for posterior inference, publicly available at: \href{https://github.com/y1jHuang/nonloc_sparse_bayes}{\texttt{y1jHuang/nonloc\_sparse\_bayes}}.

We conduct extensive simulation studies to evaluate the performance of BFMAN. The results demonstrate that our method consistently outperforms existing methods in factor recovery, sparsity detection, score estimation, and accuracy in selecting the number of latent factors.  Moreover, by modeling sparsity at the level of individual scores, BFMAN provides a more nuanced and realistic characterization of latent behavior, making the model particularly well-suited for complex, high-dimensional applications.

To further showcase the utility of our approach, we apply BFMAN to the Hispanic Community Health Study/Study of Latinos (HCHS/SOL) \citep{national2009hispanic}, a multi-center epidemiologic study designed to investigate critical components impacting the health of Hispanic/Latino populations \citep{sorlie2010design}. A key aim  of the study is the association of diet in cardiovascular disease risk factors, including diabetes, hypertension, and high cholesterol~\citep{daviglus2014cardiovascular}. Using our method, we uncover interpretable dietary patterns and their associations with these three risk factors, providing novel understanding of the diet-disease relationship.

The paper is organized as follows. Section \ref{sec:factor_analysis} introduces the BFMAN framework, the proposed mass-nonlocal prior for the factor score, and the new procedure for selecting the optimal number of factors.
Section \ref{sec:simulation} presents extensive simulation studies comparing BFMAN to standard methods. Section \ref{sec:application} applies the BFMAN model to the HCHS/SOL data. Finally, Section~\ref{sec:conclusion} includes a discussion of our findings and their implications.

\section{Bayesian mass-nonlocal factor analysis}
\label{sec:factor_analysis}

\subsection{Model and prior specification}
\label{subsec:model}
Let $\bfY \in \mathbb{R}^{n \times p}$ be the observed data matrix  where $n$ is the number of observations and $p $ is the number of variables. 
The latent factor model for each observation $i \in \{ 1, \cdots, n\}$, is given by:

\begin{align}
    {\bf y}_i = \Lambda \bfeta_i + \bfepsilon_i,
    \label{eq:model_y}
\end{align}

where $\Lambda \in \mathbb{R}^{p \times k}$ is the factor loading matrix, $\bfeta_i \in \mathbb{R}^k$ is the latent factor score vector for the $i$-th observation, where $k$ indicates the number of factors, and  $\bfepsilon_i \sim \mathcal{N}_p(0, \Sigma)$ is the idiosyncratic error matrix, with $\Sigma = \mbox{diag}(\sigma_1^2, \cdots, \sigma_p^2)$.

\begin{figure}[h]
    \centering
    \includegraphics[width=3.3in]{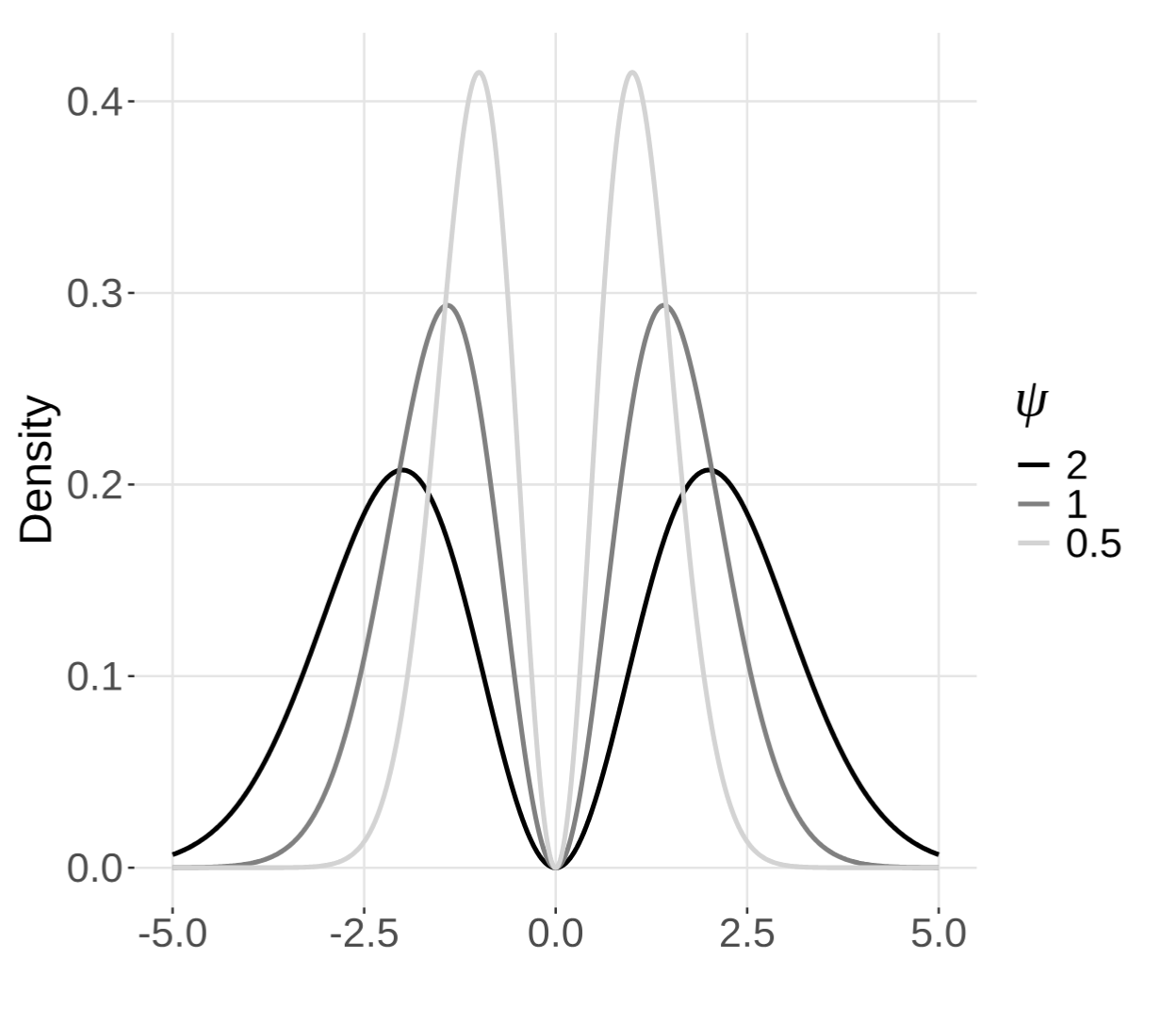}
    \caption{pMOM densities for different dispersion parameters $\psi$.  This distribution defines the slab component of our mass-nonlocal prior.}
    \label{fig:pMOM}
\end{figure}

Traditional factor models assume $\bfeta_i \sim \mathcal{N}(0, I_k)$, which may not capture the sparsity and heterogeneity in the factor scores often observed in practice \citep{tucker1971relations, distefano2019understanding}. To address this, we propose a mixture prior on the factor score $\eta_{ih}$, for each observation $i\in \{1,\dots, n\}$ and factor $h\in \{1, \dots, k\}$, that includes a Dirac distribution with mass in zero and a slab component given by a product moment (pMOM) prior \citep{johnson2010use, johnson2012bayesian}:
\begin{equation}
    \{\eta_{ih}|\theta_{h}, \phi_h \} \sim (1 - \theta_{h})\delta_0(\eta_{ih}) + \theta_{h}\mbox{pMOM}(\eta_{ih} \mid \psi_h),
    \label{eq:mass-nonlocal-prior}
\end{equation}
where $\delta_0(\cdot)$ is a Dirac measure with mass at zero, and $\mbox{pMOM} (\cdot)$ has probability density:
\begin{equation*}
p(\eta_{ih} \mid \psi_h) = \frac{1}{\sqrt{2\pi \psi^3}} \exp\left(-\frac{\eta_{ih}^2}{2\psi}\right)\eta_{ih}^2, 
\end{equation*}
with scale parameter $\psi > 0$.
This choice ensures a flexible distribution that avoids overlap with the spike in zero while preserving tails similar to a normal distribution.
Figure~\ref{fig:pMOM} illustrates the shape of the pMOM density across different values of $\psi$, highlighting its non-locality and zero-avoiding property.

The prior elicitation for the hyperparameters in the distributions in \eqref{eq:model_y} and \eqref{eq:mass-nonlocal-prior}, respectively, the variance of the idiosyncratic error and the weights of the mixture distribution, is defined as follows: 
\begin{gather}
    \sigma_j^{-2} \sim \mbox{Ga}(a_\sigma, b_\sigma) \quad \forall j\in \{1,\cdots, p\}, \notag \\
    \theta_h \sim \mbox{Beta}(a_{\theta}, b_{\theta})\quad \forall h\geq 1. \label{eq:theta} 
\end{gather}

The formulation of the mass-nonlocal prior \eqref{eq:mass-nonlocal-prior} allows us to introduce a latent variable $Z_{ih}$ for each observation $i \in \{1, \dots, n\}$ and factor $h \in \{1, \dots, k\}$, denoting whether $\eta_{ih}$ is drawn from the spike or the non-local slab with the following distribution: 
\begin{equation*}
    Z_{ih} \sim \mbox{Bern}(\theta_h),
\end{equation*}
where $\theta_h$ is the probability for the corresponding factor score $\eta_{ij}$ to follow a pMOM probability distribution, with prior \eqref{eq:theta}, such that
\begin{equation*}
    \{\eta_{ih}|Z_{ih}=1, \psi \} \sim \mbox{pMOM}(\psi) \; \mbox{ and }\;
    \{\eta_{ih}|Z_{ih}=0 \} \sim \delta_0.
\end{equation*}

For the factor loading matrix $\Lambda$, we adopt the multiplicative gamma process shrinkage (MGPS) prior  \citep{bhattacharya2011sparse}:
\begin{gather}
    \lambda_{jh}|\phi_{jh}, \tau_h \sim \mathcal{N}(0, \phi_{jh}^{-1}\tau_h^{-1}), \notag\\
    \phi_{jh} \sim \mbox{Ga}(\nu/2, \nu/2), \notag \\
    \tau_h=\prod_{l=1}^{h} \delta_l,\quad
    \delta_1 \sim \mbox{Ga}(a_1, 1),\quad \delta_l \sim \mbox{Ga}(a_2, 1),\quad l \ge 2,
    \label{eq:mgps}
\end{gather}
where $\{\tau_h\}_{h\geq 1}$ increases with $h$. This prior induces increasing shrinkage on higher-indexed columns of $\Lambda$. 
 Alternative priors such as the cumulative shrinkage prior \citep{legramanti2020bayesian} or generalized MGPS \citep{schiavon2022generalized} could also be used.

We adopt the recommended hyperparameter settings from \citet{bhattacharya2011sparse} and \citet{durante2017note}, ensuring stability and efficiency in posterior inference.

\subsection{Posterior Computation}

We develop an efficient Gibbs sampler for posterior computation, levering the conjugate prior with the exception of the pMOM distribution, which requires a Metropolis-Hasting step. 

Following the steps in the algorithm~\ref{algoritm: posterior_comp}, in each iteration $r=1,\dots, R$, we use the observed data ${\bf y}$ to update the parameters and random variables.
Let $\bflambda_j$ denote the $j$-th row of the factor loading matrix $\Lambda$, for $j \in \{1, \dots, p\}$, and $\bfeta_i$ the $i$-th row of the latent score matrix $\bfeta$, for $i \in \{1, \dots, n\}$. We indicate with ${\bf y}^{(j)}=(y_{1j}, \cdots, y_{nj})^\intercal$ the $j$ variable across all individuals. Let 
 $D_j^{-1} = \mbox{diag}(\phi_{j1}\tau_1, \cdots, \phi_{jk}\tau_{k})$ denote the diagonal prior precision matrix for
MGPS prior \eqref{eq:mgps}, and $\{\psi_j\}_{j \in \{1, \dots, k\}}$ the pMOM parameters.

Then the steps for posterior sampling are as follows:

\begin{enumerate}
    \item 
    The loading matrix entries $\{\bflambda_j\}_{j \in \{1, \dots, p\}}$ are sampled from the following posterior distribution:
        \begin{align*}
            f(\bflambda_j|\tau, \Lambda,D, \sigma_y) \sim \mathcal{N}_{k}\Big\{\left(D_j^{-1}+\sigma_j^{-2}\bfeta^\intercal \bfeta\right)^{-1}\bfeta^\intercal\sigma_j^{-2}y^{(j)}, \left(D_j^{-1}+\sigma_j^{-2}\bfeta^\intercal \bfeta\right)^{-1}\Big\}.
        \end{align*}
    
    \item The the MGPS prior introduces two parameter. First, the local shrinkage parameter $\phi_{jh}$, with the following poster distribution, for $j \in \{1, \dots, p\}$ and $h \in \{1, \dots, k\}$:
        \begin{align*}
            f(\phi_{jh}|\nu,\Lambda,\tau) \sim \text{Ga}\left(\frac{\nu+1}{2}, \frac{\nu+\tau_h\lambda_{jh}^2}{2}\right).
        \end{align*}
    \item Second, the global shrinkage parameter $\delta_h$, with posterior distribution defined as follows:
        \begin{align*}
            f(\delta_h|a,\tau,\phi,\Lambda) \sim \text{Ga}\Big\{a_h+\frac{p}{2}\left(k-h+1\right), 1+\frac{1}{2}\sum_{l=1}^{k}\tau_l^{(h)}\sum_{j=1}^{p}\phi_{jl}\lambda_{jl}^2\Big\},
        \end{align*}
        where $\tau_l^{(h)}=\prod_{t=1,t\ne h}^l \delta_t$ for $h=1,\cdots,k$.
    \item The factor score $\eta_{ih}$, conditional to the latent variable $Z_{ih}$, are sampled from: 
        \begin{align*}
            f(\eta_{ih}|Z_{ih}) = \begin{cases}
            \pi(\eta_{ih}|c,d) & \mbox{ if } Z_{ih}=1,\\
                0 & \mbox{ if } Z_{ih}=0;
            \end{cases}
        \end{align*}
    where $\pi(\eta_{ih}|-)$ indicates the following distribution:
        \begin{gather*}
            \pi(\eta_{ih}|c,d) \propto \exp\bigg\{-c\left(\eta_{ih} - \frac{d}{c}\right)^2 \bigg\} \eta_{ih}^2;\\
            \mbox{with } c = \frac{1}{2\psi} + \sum_{j=1}^p\frac{1}{2\sigma_j^2}\lambda_{jh}^2 \mbox{ and } d=\sum_{j=1}^{p}\frac{1}{2\sigma_j^2}\lambda_{jh}\left(y_{ij}-\sum_{l\neq h}^{k}\lambda_{jl}\eta_{il}\right).
        \end{gather*}
    Due to the non-conjugacy of $\pi(\eta_{ih}|c,d)$, we embedded a Metroplis-Hastings algorithm, which is implemented as follows: \\
    \begin{algorithm}[H]
    \SetAlgoLined
    \KwIn{Probability density $\pi(\eta_{ih}|c,d)$, initial state $\eta_{ih}^0$}
    \KwOut{Posterior samples from $\pi(\eta_{ih}|c,d)$}
    \For{\(m = 1\) \KwTo \(M\)}{
            Generate a random candidate $\eta_{ih}^\ast \sim \mathcal{N}(\mu=\eta_{ih}^{m-1}, \sigma)$;\\
            Calculate acceptance probability $r = \exp \left(\log\pi(\eta_{ih}^\ast|c,d) - \log\pi(\eta_{ih}^{m-1}|c,d) \right)$;\\
            Accept or reject:\\
                \Indp$\alpha = \min(1, r)$,\\ $Z = \mbox{Bern}(\alpha)$,\\
                $\eta_{ih}^m = Z \eta_{ih}^\ast + (1-Z)\eta_{ih}^{m-1}$.
    }
    \caption{Metroplis-Hastings Algorithm} 
    \label{algoritm: posterior_comp_MH}
    \end{algorithm}
    \item 
    Sample latent variable $Z_{ih}$, for $i \in \{1, \dots, n\}$ and $h \in \{1, \dots, k\}$,  from a Bernulli distribution such that
        \begin{align*}
            \mbox{Pr}(Z_{ih}=0|-) &= \frac{f(Z_{ih}=0)f(\bfyi|Z_{ih}=0,-)}{f(Z_{ih}=1)f(\bfyi|Z_{ih}=1,-)+f(Z_{ih}=0)f(\bfyi|Z_{ih}=0,-)} \\
            &= \frac{f(Z_{ih}=0)}{f(Z_{ih}=0)+f(Z_{ih}=1)T},
        \end{align*}
        where $T = K \sqrt{2\pi}H^{-\frac{1}{2}}(H^{-1}+M^2)$, $H=\frac{1}{\psi} +\bflambda_h^\intercal\Sigma^{-1}\bflambda_h$, $M = \frac{1}{H}(\bfyi - \Lambda_{(-h)} \bfeta_{i(-h)})\Sigma^{-1}\bflambda_h$ and $K=2\pi \psi^{-\frac{3}{2}}\exp\{\frac{1}{2}HM^2\}$. The $\Lambda_{(-h)}$ represents $p\times(k-1)$ matrix with $h$th column dropped, and $\bfeta_{i(-h)}$ denotes $k-1$ vector with $\eta_{ih}$ entry deleted.
    \item Sample the probability parameter $\theta_h$, for each factor $h \in \{1, \dots, k\}$, from:
        \begin{align*}
            f(\theta_h |{\bf Z}_{h}, a_1, b_1) = \mbox{Beta}\left(\sum_i^nZ_{ih}+a_1, n-\sum_i^nZ_{ih}+b_1\right),
        \end{align*}
        where $a_1$ and $b_1$ are the hyperparameters.
    \item The residual variance $\sigma_j^{2}$, for $j \in \{1, \dots, p\}$, is drawn from the posterior distribution:
        \begin{align*}
            f(\sigma_j^{-2}|a_{\sigma}, b_{\sigma},y, \Lambda, \bfeta) = \text{Ga}\Big\{a_{\sigma}+\frac{n}{2}, b_{\sigma}+\frac{1}{2}\sum_{i=1}^{n}\left(y_ij-\bflambda_j^\intercal\bfeta_i\right)^2\Big\}.
        \end{align*}
\end{enumerate}

 \begin{algorithm}[H]
     \SetAlgoLined
     \KwIn{Outcome matrix $\bf Y$}
     \KwOut{Posterior distribution of each parameter}
     \For{\(r = 1\) \KwTo \(R\)}{
             Sample factor loading $\bflambda_j$ for $j \in \{1, \dots, p\}$;\\
             Sample $\phi_{jh}$, for $j \in \{1, \dots, p\}$ and $h \in \{1, \dots, k\}$;\\
             Sample $\delta_h$ for $h \in \{1, \dots, k\}$;\\
             Sample factor score $\eta_{ih}$ given the latent variable $Z_{ih}$, for $i \in \{1, \dots, n\}$ $h \in \{1, \dots, k\}$;\\
             Sample the latent variable $Z_{ih}$, for $i \in \{1, \dots, n\}$ and $h \in \{1, \dots, k\}$;\\
             Sample the $\theta_h |{\bf Z}_{h}$, for $h \in \{1, \dots, k\}$;\\
             Sample the residual variance $\sigma_j^2$ for $j \in \{1, \dots, p\}$.
     }
    \caption{Posterior computation.} 
     \label{algoritm: posterior_comp}
 \end{algorithm}

\subsection{Model Identification}

Latent factor models are non-identifiable due  to their invariance under orthogonal transformations. Specifically, for any orthogonal matrix $Q \in \mathbb{R}^{k \times k}$, the latent structure can be equivalently expressed as $\Lambda^* = Q \Lambda$ and $\bfeta^* = \bfeta Q^\top$. As a result, the model can be rewritten as:
$
{\bf y}_i = \Lambda^* \bfeta^*_i + \bfepsilon_i$,
highlighting the rotational ambiguity in the factorization of the latent space.

To address this non-identifiability issue, several approaches have been proposed in the literature. Standard solutions include imposing structural constraints on the loading matrix $\Lambda$. For instance, \citet{lopes2004} enforce a lower-triangular structure with strictly positive diagonal entries to ensure uniqueness. Classical rotation methods such as the varimax criterion  \citep{kaiser1960application},  aim to improve interpretability by maximizing the variance of squared loadings post-rotation. More recent approaches, including the parameter expansion framework of \citet{rovckova2016fast}, extend this idea by proposing EM-based optimization schemes that seek sparse, rotated loading matrices. Their approach mitigates the risk of local optima by expanding the parameter space, enabling greater flexibility in the estimation process. This approach was then followed by
\citet{avalos2022heterogeneous}, addressing identifiability solely through adding sparsity in the factor loadings via a non-local mass prior.
This is further corroborated by the recent paper of \citet{fruhwirth2024sparse}, the generalized lower-triangular (GLT) decomposition. The decomposition introduces a structure of $\Lambda$ that satisfies the following condition: for each column $h \in \{1,\cdots, k\}$, let $l_h$ denote the row index of its leading non-zero entry. Then the indices $l_1,l_2, \cdots, l_k$ must be in ascending order, i.e. $l_1 < l_2 < \cdots, l_k$, while the leading entries $\Lambda_{l_h, h} > 0$.

While previous work has primarily addressed non-identifiability by constraining the loading matrix, our contribution lies in a different and novel direction by enforcing identifiability on the factor score matrix $\bfeta$, instead. This represents a key innovation of our model, as, to the best of our knowledge, no existing work has considered identifiability from the perspective of the factor score matrix. Specifically, we extend the GLT decomposition \citep{fruhwirth2024sparse} to the factor score matrix, $\bfeta$. By applying the GLT constraints to the scores, we simultaneously achieve identifiability and parsimony with fewer assumptions in the latent space, both $\Lambda$ and $\bfeta$. Under this structure, the only possible rotation in the score is the identity matrix, eliminating rotational ambiguity in a natural and interpretable way \citep{fruhwirth2024sparse}.

Furthermore, the sparsity induced by the GLT structure not only resolves identifiability, but also facilitate inference on the number of factors.

\subsection{Choosing number of factors} 
\label{subsec:nfac}

Determining the optimal number of factors is a critical challenge in factor analysis. The objective is to retain a small number of factors that captures the underlying structure of the data without introducing redundancy. Traditional methods for factor number selection  typically fall into two broad categories: threshold-based methods and model selection via information criteria.

Threshold-based methods, such as Kaiser’s criterion \citep{kaiser1960application} or scree plots \citep{castello2016association}, rely on thresholds, such as retaining factors with corresponding eigenvalues greater than one, based on Principal Component Analysis (PCA). Although computationally efficient, these approaches are sensitive to the specific structure and variability of the dataset, often resulting in inconsistent or unstable estimates.

Information-theoretic criteria, including the Bayesian Information Criterion (BIC) or the Akaike Information Criterion (AIC) \citep{kp1998model, preacher2012problem}, offer a more principled alternative by comparing models with different numbers of factors \citep{bai2002determining}. However, these methods are computationally intensive, especially in high-dimensional settings, as each model configuration must be fit and evaluated independently. 

Positioned between heuristic methods and information-theoretic criteria, \citet{bhattacharya2011sparse} introduces an adaptive shrinkage approach that starts with a large number of factors and iteratively prunes redundant ones by shrinking the columns of the loading matrix. This strategy still relies on thresholding decisions—defining when a column is "close enough" to zero to be removed—and is sensitive to the tuning of shrinkage hyperparameters \citep{durante2017note}.

To address these challenges, we introduce a novel method for estimating the number of factors by focusing on sparsity and identifiability in the factor score matrix, not the loadings. Our approach shifts the focus to the factor score matrix $\bfeta$, leveraging the sparsity-inducing mass-nonlocal prior introduced in \eqref{eq:mass-nonlocal-prior},  motivated by variable selection framework.

Under this formulation, begin with a conservative upper bound $K = 5 \log(p)$ \citep{bhattacharya2011sparse} following \citet{bhattacharya2011sparse}, and successively use posterior inference to determine the important factors while discarding the irrelevant ones.  Specifically, the posterior estimation of the factor score $\bfeta$ provides insights into which entries can be considered effectively zero and which deviate significantly from zero.  Columns where a high proportion (e.g., $\geq 80\%$) of entries are exactly zero are considered not important and discarded. This procedure is guided by the posterior distribution over the latent indicators $Z_{ih}$, which directly reflect whether an individual expresses a given factor.

Our proposed sparsity-inducing prior for the score matrix $\bfeta$ allows us to automatically estimate the number of relevant factors, without relying on arbitrary thresholds for the loadings or model comparisons.
This strategy shifts the identifiability constraint from the loading matrix to the score matrix, reducing the need for strong structural assumptions and providing a clear and interpretable mechanism for factor selection. Moreover, it results in a computationally efficient and flexible method that avoids overestimation and adapts naturally to the structure of the data.


\section{Simulation study}
\label{sec:simulation}

We conduct extensive simulation experiments to evaluate the performance of our BFMAN in recovering the sparse structure of the factor score matrix $\bfeta$ and the factor loading matrix $\Lambda$. A particular focus is placed on the model’s ability to correctly identify and impute the zeroes entries in $\bfeta$. 
To benchmark the performance,  we compare our method with the MGPS factor model by \citet{bhattacharya2011sparse}.

We construct four distinct simulation scenarios with varying levels of complexity in the data generation process, focusing on different sparsity schemes in the factor scores. The data generation process differs across scenarios in terms of sample size $n$, number of observed variables $p$, number of factors $k$, and the sparsity of the factors score induced by $\{\theta_h\}_{h=1}^{k}$. 
Table~\ref{tab:simulation_spec} summarizes the scenario-specific parameters, while Table~\ref{tab:simulation_common} describes the data-generating process used across all scenarios.

Specifically, Scenario 1 represents a setup with a small $n$ and $p$. Scenario 2 retains the same dimensions as Scenario 1 but introduces heterogeneity in the factor scores sparsity, with increasing probabilities of zeros across factors columns.
Scenarios 3 reflects a high-dimensional setting where the number of variables exceeds the sample size, i.e., $p >> n$. Finally,  Scenario 4 mimics our real data nutritional application analyzed in Section \ref{sec:application}. 
Each scenario is replicated 50 times.

\begin{table}[]
\caption{Scenario-specific parameters used to generate the simulation experiments.}
    \centering
    \begin{tabular}{c|cccc}
    \multicolumn{1}{c}{  } \\
         & Scenario 1 & Scenario 2 & Scenario 3 & Scenario 4 \\
         \hline
        $n$ & $100$ & $100$ & $30$ & $3 000$  \\
        $p$ & $20$ & $20$ & $60$ & $60$\\
        $k$ & $3$ & $3$ & $5$ & $6$ \\
        $\{\theta_h\}_{h=1}^{k}$ & $0.4\; \forall h$ & $\{0.8,0.6,0.4\}$ & $\{0.9, 0.8,0.7,0.6,0.5\}$ & $\{0.8, 0.7, 0.6, 0.5, 0.4, 0.3\}$ 
    \end{tabular}
    \label{tab:simulation_spec}
\end{table}

\begin{table}
\caption{Data generating mechanism across the scenarios, for $i \in \{1, \dots,n\}$ and $h \in \{1, \dots, k\}$. 
}
    \centering
    \begin{tabular}{c}
    \multicolumn{1}{c}{  } \\
    \hline
    $Z_{ih} \sim \mbox{Bern}(\theta_{h}),$\\
    $\eta_{ih} \sim (1 - Z_{ih})\delta_0 + Z_{ih}\mbox{pMOM}(\psi=0.5),$ \\
    $\bflambda_{h} \sim \Norm_p(0,\mathbb{I}),$  \\
    $\bfepsilon_i \sim \Norm_p(0, \Sigma)$ with $\Sigma=\mbox{diag}(\sigma^2_1, \dots, \sigma^2_p),$\\
    $\sigma^2_j \sim \mbox{Unif}(0,1)\; \forall j\in \{1, \dots, p\},$ \\
    $\bfY_i = \Lambda \bfeta_i + \bfepsilon_i$. \\
    \hline
    \end{tabular}
    \label{tab:simulation_common}
\end{table}

To evaluate the ability to recover the true latent structure---factor loading $\Lambda$ and factor score $\bfeta$---we compute the RV coefficient \citep{robert1976unifying}. The RV coefficient compares the estimated and true structures, returning a value between from 0 (no similarity) to 1 (higher similarity).

\begin{figure}[h!]
    \centering
    \includegraphics[width=5in]{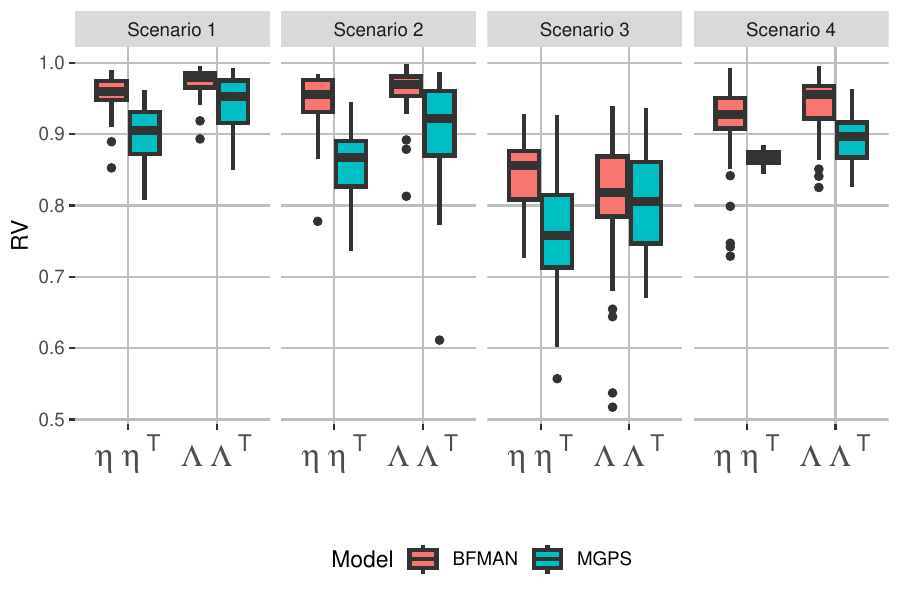}
    \caption{Results comparison: RV coefficient for $\bfeta\bfeta^T$ and $\Lambda\Lambda^T$ estimated with our BFMAN model (in red) and MGPS model (in blue) across the four simulated scenarios.}
    \label{fig:results-RV}
\end{figure}

Figure~\ref{fig:results-RV} reports the RV results across all the scenarios. The 
BFMAN model consistently achieves high RV coefficients, demonstrating excellent recovery of the latent structure. In Scenarios 1 and 2, the RV values exceed $0.95$ for both $\Lambda\Lambda^T$ and $\bfeta\bfeta^T$, indicating near-perfect recovery and overperforming MGPS model. 
Even in the more challenging scenarios, which closely mimic real-world data complexities, our model consistently outperforms the MGPS model, while the RV index values remain close to 1. In Scenario 3, where $p > > n$, the RV coefficients are respectively $0.85$ for $\bfeta\bfeta^T$ and $0.8$ for $\Lambda\Lambda^T$. In Scenario 4, with a large sample size and a high-dimensional multivariate variable, both matrices achieve an RV index greater than $0.9$. These results highlight the superior ability of the proposed model in accurately recovering the underlying data structure across different levels of sparsity and dimensionality.

\begin{figure}[h!]
    \centering
    \includegraphics[width=5in]{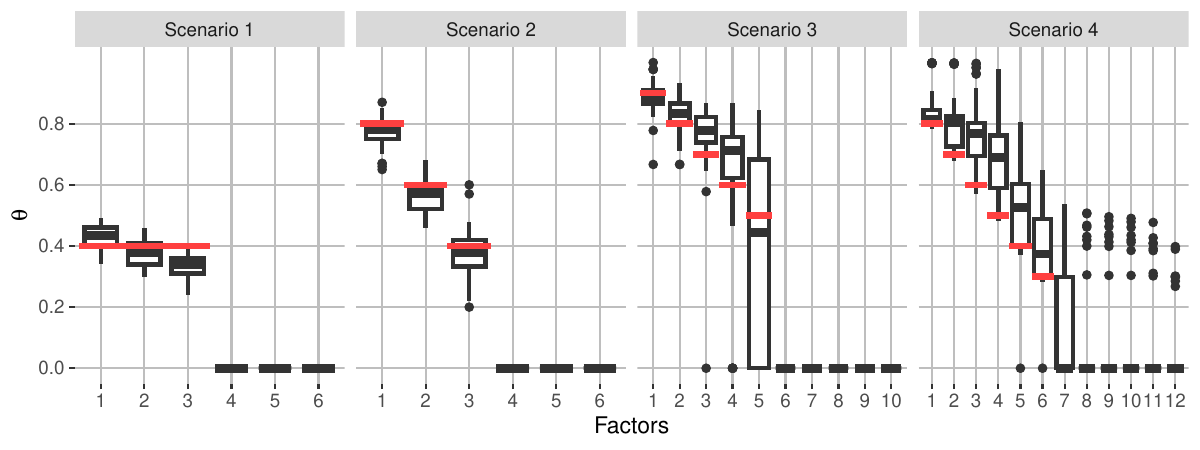}
    \caption{Results for BFMAN model. Distribution, over the 50 replicates, of the estimated probability of nonzero entries in the factor score matrix $\theta$ for each factor and for each of the four simulated scenario. The red line indicated the true value. For the factors where the red line is not reported, the true value is zero, indicating the absence of that factor in the data generating process.}
    \label{fig:results-theta}

\end{figure}

As indicated in the previous section, the key strength of our model is its ability to identify the sparsity of the factor score matrix and to exploit it to determine the number of factors. Therefore, Figure \ref{fig:results-theta} illustrates the distribution of the estimated probabilities, $\hat{\theta}$, of non-zero entries in $\bfeta$ for each factor across the 50 replicates. 
The estimates match closely the true simlated value (in red), falling within the interquartile range, demonstrating accurate recovery of the sparsity structure. 
Our method not only accurately estimates the proportion of nonzero entries in the factor score matrix but also correctly assigns $\theta \approx 0$ to the additional factors $k^\ast = K - k$ that are used to estimated the model but are not part of the data-generating process, where $K$ represents the upper bound used in model estimation.


\begin{figure}[h!]    
    \centering
    \includegraphics[width=5in]{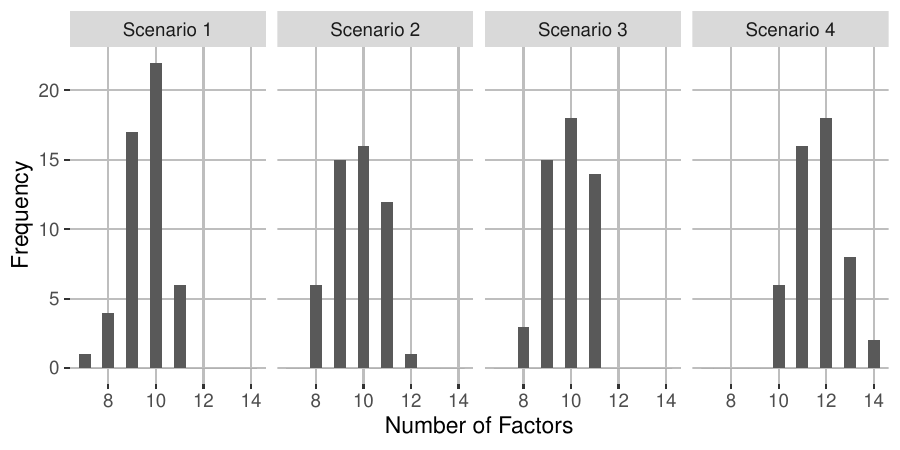}
    \caption{Results for MGPS model. Distribution, over the 50 replicates, of the probability of nonzero entries in the factor score matrix $\theta$ estimated for each factor and for each of the four simulated scenario. }
    \label{fig:results-estK_MGPS}
\end{figure}

In contrast, the MGPS model exhibits a tendency to overestimate the number of factors, as illustrated in Figure~\ref{fig:results-estK_MGPS}. This suggests that MGPS may require further tuning of its penalty parameters to better control shrinkage and avoid selecting spurious factors—especially in settings with small sample sizes and low dimensions, where overestimation is more likely.

These findings further corroborate the superior performance of our model in accurately selecting the true number of latent factors while preserving interpretability and sparsity.
Furthermore, our factor selection procedure—based on discarding columns in $\bfeta$ that are entirely or mostly zero—shows crucial advantages.  In Scenarios 1 and 2, all true factors are correctly retained across all replicates. In Scenario 3, the overall identification remained accurate; however, our model occasionally underestimated the number of factors, particularly for Factor 5, which had the highest level of sparsity. Although Scenario 4 slightly overestimates the number of factors in a few replicates, MGPS consistently shows a much greater overestimation,  as shown in Figure \ref{fig:results-estK_MGPS}). In all four scenarios, the MGPS model tends to estimate nearly twice the true number of simulated factors.
These results further corroborate the performance of our model in accurately selecting the true number of latent factors while preserving interpretability and sparsity.

\section{Nutritional data and cardiovascular diseases}
\label{sec:application}

In this section, we apply our proposed model to the Hispanic Community Health Study/Study of Latinos (HCHS/SOL), a large-scale, multi-site cohort designed to investigate the relationship between diet and cardiovascular risk factors in Hispanic/Latino population.  The study includes 14,002 adults aged 18-74 years from four U.S. cities (Bronx, Chicago, Miami, and San Diego), recruited using a stratified two-stage probability sampling design as detailed in \citet{lavange2010sample}.

From the original dataset, we exclude individuals who are on relevant medication therapy, have missing data, and/or present unreliable dietary questionnaires (e.g., extreme energy intakes, negative values for nutrient or food intake, or poor quality reported by interviewers) as described in \citet{de2023multi}. The resulting dataset includes 2,273 subjects and 53 nutrients. All nutrient values are log-transformed and standardize prior to analysis.

We first estimate the latent dimensionality using the strategy outlined in Section~\ref{subsec:nfac} starting with $K=5\log(p)$, i.e., $K=12$. Then we discard factors in which at least 85\% of the entries in the score matrix are zeros, yielding a final model with 6 factors. Then we rerun the factor analysis setting $k = 6$ to obtain the factor loading and the score matrix.

\begin{figure}[h!]
    \centering
    \includegraphics[width=\textwidth]{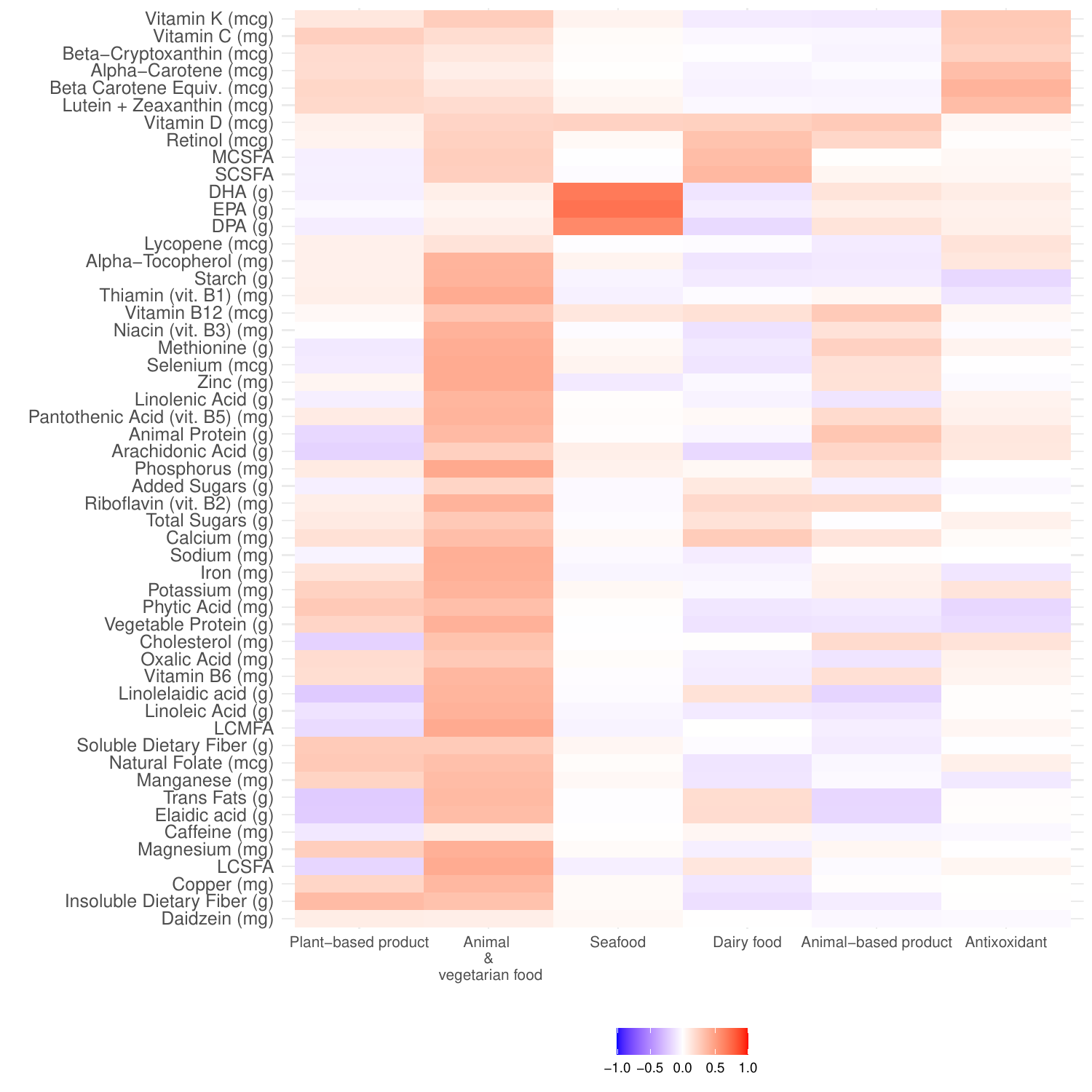}
    \caption{Heatmap of the factor loadings in the HCHS/SOL estimated with BFMAN.}
    \label{fig:factor_score_nutrients}
\end{figure}

We then proceed to interpret the estimated factor loading matrix, reported in Figure \ref{fig:factor_score_nutrients}. Following nutritional literature, we name each factor based on important loadings, i.e.$\lambda_{ih}\geq 0.3$. The first factor, namely \textit{plant-based products}, is characterized by high loadings on insoluble and soluble dietary fiber, magnesium, natural folate, and phytic acid. 
The second factor, labeled \textit{animal and vegeterian food}, reflects a complex, nutrient-dense pattern that incorporates a wide array of nutrients from both plant and animal sources. It includes various proteins, essential fatty acids (such as linoleic, linolenic, LCSFA, and LCMFA), cholesterol, trans fats, a wide range of minerals (including calcium, iron, zinc, and magnesium), and several vitamins (particularly the B-complex and vitamin E).
The third factor, namely the \textit{seafood} pattern, is defined by high factor loadings of omega-3 fatty acids, such as eicosapentaenoic acid (EPA), docosapentaenoic acid (DPA), and docosahexaenoic acid (DHA).
The fourth factor, labeled \textit{dairy products}, shows significant contributions from short- and medium-chain saturated fatty acids (SCSFA and MCSFA), calcium, and retinol.
The fifth factor, representing \textit{animal products},is driven by high loadings on animal protein, vitamin B12, and vitamin D.
Finally, the sixth factor, named \textit{antioxidant products} factor, includes lutein and zeaxanthin, beta carotene, alpha-carotene, and vitamin C, that highlight the antioxidant content of the diet.

Then, we proceed on estimating their association with key cardiovascular risk factors: diabetes, high cholesterol, and hypertension. We fit a Bayesian logistic regression for each outcome, including confounders such as energy, physical activity, depressive symptoms (CESD score), ethnicity, gender, employment, years as US residency, marital status, income, education,  alcohol and tobacco use.

\begin{figure}[h!]
    \centering
    \includegraphics[width=\textwidth]{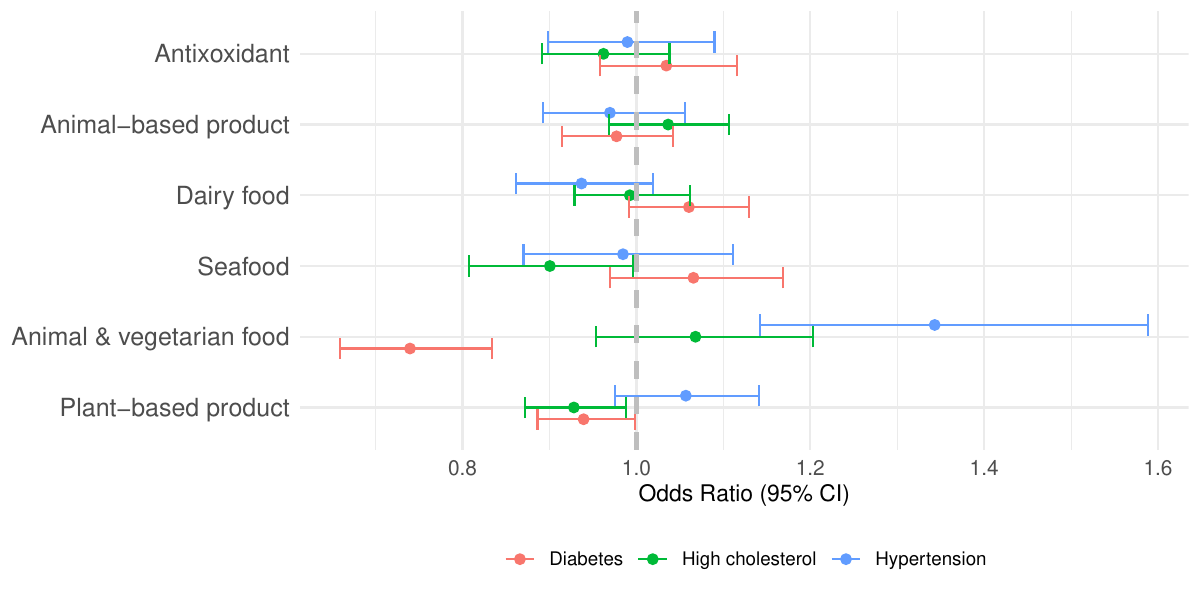}
    \caption{Odds ratio and their corresponding 95\% credible intervals for each dietary pattern for the three CVD risk factors: diabetes, hypertension, and high cholesterol.}
    \label{fig:OR}
\end{figure}

The results reported in Figure \ref{fig:OR} show that the plant-based product pattern is inversely associated with the risk of diabetes and high cholesterol,  aligning with previous evidence on the protective effects of vegetarian diets against cardiometabolic diseases \citep{kahleova2018vegetarian}. 
The animal and vegetarian pattern has a double trend: it is positively associated with hypertension but inversely associated with diabetes.  This factor includes both beneficial components such as fiber, linoleic acid, and plant proteins, and potentially adverse components like cholesterol, animal protein, and saturated fats, contributing to these mixed associations. 
Finally, the pattern of seafood consumption is inversely associated with the risk of high cholesterol, supporting previous evidence that seafood consumption is protective against cardiovascular risk factors \citep{aadland2015lean}.

This application illustrates the practical advantages of our method in an epidemiological setting. By inducing sparsity in the factor score matrix, our model not only automatically estimates the number of meaningful latent dietary patterns but also enhances interpretability, allowing for a clearer identification of associations between diet and disease. These features make our approach particularly well suited to uncovering actionable insights in complex, high-dimensional health data.

\section{Discussion}
\label{sec:conclusion}

In this paper, we introduce a novel Bayesian factor model---the BFMAN---that shifts the focus from the commonly studied factor loadings to the factor scores. By incorporating a non-local mass prior on the factor scores, our BFMAN effectively captures individual-level heterogeneity and enforces sparsity in a principled manner. This leads to a richer and more realistic representation of how each subject contributes to latent structures, as demonstrated in our nutritional epidemiology application, where individual adherence to specific dietary patterns varied substantially.

Moreover, the sparsity plays a key role in (i) the methodological aspect of defining a robust and novel approach to determining the optimal number of factors, and (ii) the real-world application,  allowing for a clearer interpretation of factor scores and highlighting which individuals meaningfully engage with certain latent patterns. In the nutritional setting, zero entries in the factor score matrix help identify individuals who do not follow particular dietary behaviors, thereby reducing noise and improving interpretability.

Our simulation results show that BFMAN consistently outperforms the widely used MGPS model. Across all scenarios, BFMAN achieve higher RV coefficients---indicating superior recovery of both the factor score and loading matrices---and provide more accurate estimation of the true number of latent factors. These findings reinforce the model’s reliability and support its application in complex, high-dimensional settings.

In the real-world analysis of the HCHS/SOL study, BFMAN identified six interpretable dietary patterns, including plant-based foods, animal products, seafood, dairy products, antioxidants, and a nutrient-dense mixed pattern. The model revealed that only individuals with elevated consumption of processed foods showed a significantly increased probability of developing hypercholesterolemia. By leveraging sparsity in the factor scores, we were able to determine not only the most influential dietary patterns, but also the individuals who truly adhered to them, providing a clearer link between diet and health outcomes.



Our results underscore the central role of factor scores in both methodological innovation and real-world interpretation. While much of the existing literature has focused on imposing structure on the loadings, our work highlights how priors on the scores can yield powerful advantages. A related contribution by \citet{bortolato2024adaptive} introduces adaptive shrinkage priors on factor scores for multi-study settings, further validating the relevance of this direction. While different prior formulations may be suited to different applications, we believe the mass-nonlocal prior introduced here provides a flexible and interpretable foundation for modeling sparsity and heterogeneity in latent factor models.

Several extensions and generalization can be applied to the model. These include adapting BFMAN to dynamic or longitudinal settings, incorporating structured covariates into the prior on scores, and exploring alternative prior formulations for specific domains. More broadly, we hope this work inspires renewed attention on the modeling of factor scores, which hold rich and underutilized potential for inference and discovery across scientific disciplines.

\bigskip


\begin{funding}
RDV was supported by the US National Institutes of Health, NIGMS/NIH COBRE CBHD P20GM109035. This Manuscript was prepared using HCHSSOL Research
Materials obtained from the NHLBI Biologic Specimen and Data Repository Information Coordinating Center and
does not necessarily reflect the opinions or views of the HCHSSOL or the NHLBI.
\end{funding}


\bibliographystyle{ba}
\bibliography{ref}

\end{document}